\documentclass[11pt]{JHEP3}

\usepackage{amsmath,amsfonts,amssymb,epsfig}

\usepackage{cancel}
\usepackage{cite}
\usepackage{verbatim}

\newcommand{\A}{\A_{\mu}^a}

\let\a=\alpha         
    \let\h=\eta     
        
           \let\p=\pi      
\let\s=\sigma  
     
     \let\y=\psi    
\def\nn{\nonumber}
\newcommand{\be}{\begin{equation}}
\newcommand{\ee}{\end{equation}}
\newcommand{\bea}{\begin{eqnarray}}
\newcommand{\eea}{\end{eqnarray}}
\newcommand{\ba}{\begin{array}}
\newcommand{\ea}{\end{array}}

\title{One-loop adjoint masses for branes at non-supersymmetric angles.}
\author{
P.~Anastasopoulos$^{1}$\footnote{pascal@hep.itp.tuwien.ac.at},~
I.~Antoniadis$^2$\footnote{ignatios.antoniadis@cern.ch}~\footnote{On leave from CPHT (UMR CNRS 7644) Ecole Polytechnique, F-91128 Palaiseau}
K.~Benakli$^3$ \footnote{kbenakli@lpthe.jussieu.fr}~
M.~D.~Goodsell$^2$ \footnote{mark.goodsell@cern.at}~
A.~Vichi$^4$ \footnote{alessandro.vichi@epfl.ch}\\
$^1$ Technische Univ. Wien Inst. f\"ur Theoretische Physik, A-1040 Vienna, Austria\\
$^2$ Department of Physics, CERN Theory Division, CH-1211, Geneva 23, Switzerland\\
$^3$ Lab de Physique Th\'eorique et Hautes Energies, CNRS, UPMC Univ Paris 06, France\\
$^4$ Lawrence Berkeley National Laboratory, Physics Div, Berkeley, CA 94720-8153, USA\\
}

\date{}

\maketitle
\abstract{
This proceeding is based on arXiv:1105.0591 [hep-th] where we consider breaking of supersymmetry in intersecting D-brane configurations by slight deviation of the angles from their supersymmetric values. 
We compute the masses generated by radiative corrections for the adjoint scalars on the brane world-volumes. 
In the open string channel, the string two-point function receives contributions only from 
the infrared limits of $N\approx 4$ and $N\approx 2$ supersymmetric configurations, via messengers and their Kaluza-Klein excitations, and leads at leading order to tachyonic directions.}

 \clearpage

\begin{document}

\section{Introduction, motivation and conclusions}

Vacuum configurations with open unoriented strings have attracted a lot of attention in the past few years for their remarkable phenomenological properties \cite{Blumenhagen:2005mu, Blumenhagen:2006ci, Marchesano:2007de, Bianchi:2009va}.
One of the peculiar features is the possibility of accommodating large extra dimension and a low string tension of a few TeV, making possible the observation of stringy effects at future colliders \cite{Dudas:1999gz, Accomando:1999sj, Cullen:2000ef, Kiritsis:2002aj, Burikham:2004su, Chialva:2005gt, Bianchi:2006nf, Anchordoqui:2007da, Anastasopoulos:2008jt, Anchordoqui:2008ac, Dong:2010jt, Lust:2008qc, Lust:2009pz, Anchordoqui:2009mm, Feng:2010yx, Anastasopoulos:2011hj}. Scenarios of these kinds can be easily realized in string perturbation theory in terms of intersecting or magnetized D-branes.

One of the most interesting problems in this framework is the realization of configurations which describe softly broken supersymmetric low energy effective field theories.
Supersymmetry breaking can be easily achieved by introducing a magnetic field which, due to the different couplings with the spins, induces a mass splitting between fermions with different chiralities and with bosons \cite{Bachas:1995ik, Angelantonj:2000hi}. The same splitting can be mapped upon T-duality into branes intersecting at angles \cite{Berkooz:1996km, Blumenhagen:2000wh}. 

A supersymmetric vacuum can be obtained through a specific choice of intersection angles between D-branes. 
Then, a breaking of supersymmetry with a size parametrically smaller than the string scale can be obtained by choosing the angles slightly away from their supersymmetric values \cite{Antoniadis:2006eb, Antoniadis:2005em, Antoniadis:1997mm}. 
Supersymmetry is broken at tree-level for strings stretched between branes that intersect at non-supersymmetric angles. The breaking is communicated to the other states living on the brane world-volume through radiative corrections.

In this proceeding, which is based on arXiv:1110.5359 [hep-th] \cite{Anastasopoulos:2011kr}, we will perform an explicit computation of such effects. We will be particularly interested in the induced masses for the adjoint representations of the gauge group. This mechanism generates for instance one-loop Dirac gaugino masses, but some adjoint scalars tend to become tachyonic in the effective field theory. 
Understanding the moduli-dependance of the adjoint masses we will be able to build using this technique interesting viable models of supersymmetry breaking.

We will perform the string computation in the case of toroidal compactifications (with or without orientifold and orbifold projections) as the world-sheet description by free fields allows the straightforward use of conformal field theory techniques. 
The results depend on the number of supersymmetries that are originally preserved by the brane intersections before having the small shift in angles that induces supersymmetry breaking:
\begin{itemize}
\item The mass corrections vanish for an originally $N=1$ sector with non-vanishing intersection angles in the three tori (written as $N \approx 1$). This is due to the absence of couplings between the messengers and  scalars in adjoint representations at the one-loop level. 
\item The $N\approx2$ and $N\approx4$, one can derive the one-loop effective potential and read from there the masses of the adjoint representations. 
At leading order, the obtained mass matrix is traceless, and signals the presence of a tachyonic direction.
\end{itemize}

The string computation gives in addition a tree-level closed string divergence in the ultraviolet limit of the open string channel. 
It is shown in \cite{Anastasopoulos:2011kr} that this is actually a reducible contribution, matching the expectations from supergravity in the presence of NS-NS tadpoles through the emission of a  massless  dilaton and internal metric moduli. These results are expected to be drastically modified when taking moduli stabilization into account, causing a shift in the vacuum of the theory and cancellation of the tadpoles.

Beyond expected field theory contributions, it is interesting to find that there is no extra contribution (at leading order in the supersymmetry breaking parameter expansion) from the massive string states due to the form of the correlation functions and the boundary conditions involved in the computation of the amplitude, a feature that needed an explicit check by writing down the two-point correlation functions.

\section{D-brane setup}

\begin{figure}[h]
\begin{center}
\epsfig{file=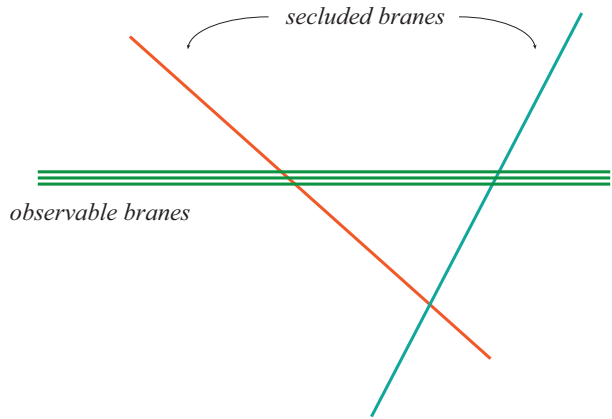,width=8cm}
\caption{Our D-brane setup. The observable branes represent a supersymmetric version of the Standard Model. The secluded branes intersect in not supersymmetric angles with the observable sector breaking supersymmetry which is communicated to the observable sector via messenger strings aka strings stretched between the observable and secluded sector. Notice that this figure is just illustrative.}
\label{Motivation_Fig}
\end{center}
\end{figure}

Our configuration contains an observable D-brane sector where our world is located (i.e. a supersymmetric version of the Standard Model). In addition, there are some secluded branes which intersect in non supersymmetric angles with the observable sector. Supersymmetry breaking will be communicated to the observable sector via strings at the intersections (see fig \ref{Motivation_Fig}) \cite{Antoniadis:2006eb, Antoniadis:2005em, Anastasopoulos:2011kr}. 

In order to perform our computations we consider toroidal compactifications of Type IIA with two D6-branes $a,~b$ in: $M_4\times T^2_1 \times T^2_2\times T^2_3$.
We assume non-SUSY configuration: 
\bea
\theta^{ab}_1 + \theta^{ab}_2 + \theta^{ab}_3 = \epsilon \approx 0
\eea
where $\theta_{ab}^i$ denotes the intersection angle of the two branes at the $i$th torus.

\begin{figure}[h]
\begin{center}
\epsfig{file=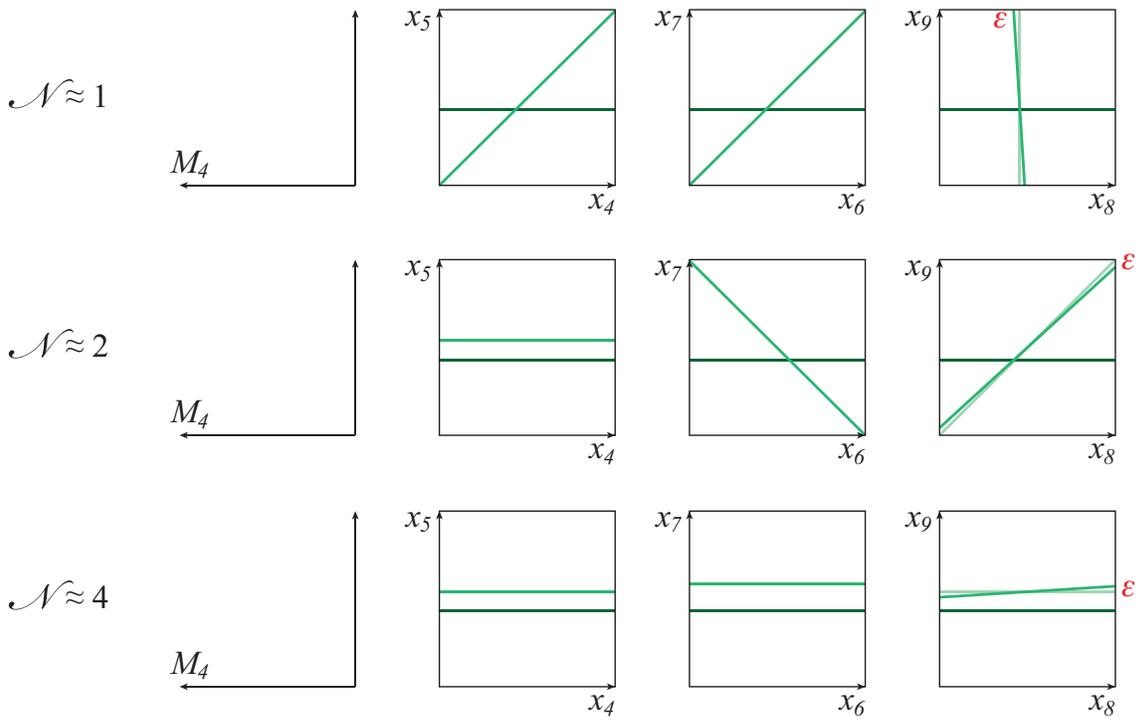 ,width=15cm}
\caption{We consider D-branes intersecting at not supersymmetric angles. At ${\cal N}  \approx 1$, branes intersect at all tori. At ${\cal N}  \approx 2,~4$ there are parallel directions at one, two tori respectively. 
Notice that this figures are again illustrative. Thus, we omit the numerous wrapping of the distorted brane for simplicity. Another way to introduce the $\epsilon$ is to distort one of the tori (the third one in this case) by changing the modulus of the torus ($\tau_3$ of the third torus). This would not introduce multiple wrappings of the brane and is closer to what we have in mind in this work.}
\label{Branes_FT}
\end{center}
\end{figure}

Different brane configurations preserve different amount of SUSY (figure \ref{Branes_FT}):
\begin{itemize}
\item At ${\cal N}  \approx 1$, branes intersect at all tori.
\item At ${\cal N}  \approx 2$ the branes are parallel in the first torus (aka $\theta^1_{ab}=0$) and intersect at non-sypersymmetric angles in the other two.
\item At ${\cal N}  \approx 4$ the branes are parallel in the first and second torus ($\theta^1_{ab}=\theta^2_{ab}=0$) and intersect with a small angle in the third torus $\theta^3_{ab}=\epsilon$.
\end{itemize}
In this framework, we will calculate the 1-loop mass of the adjoint scalars.

\section{Radiative masses for adjoint scalars}

As we mention above we will focus on radiative corrections to masses of the adjoint scalars. There are two different kind of scalars and we will calculate their masses using different techniques:
\begin{itemize}

\item Adjoint scalars in non-parallel directions. We will evaluate the 1-loop mass of such fields by the standard conformal field theory method, by inserting vertex operators (VOs) at the boundaries of the corresponding surfaces. This method is the most general and could be performed in intersecting and parallel directions.

\item Adjoint scalars in parallel directions. We will compute the partition function in the presence of brane-displacement and we will calculate the radiative corrections to the mass by taking derivatives of the displacements.
This method can be performed only on parallel directions, but it is much easier.

\end{itemize}

\subsection{Non-parallel directions by the standard amplitude method}

\begin{figure}[t]
\begin{center}
\epsfig{file=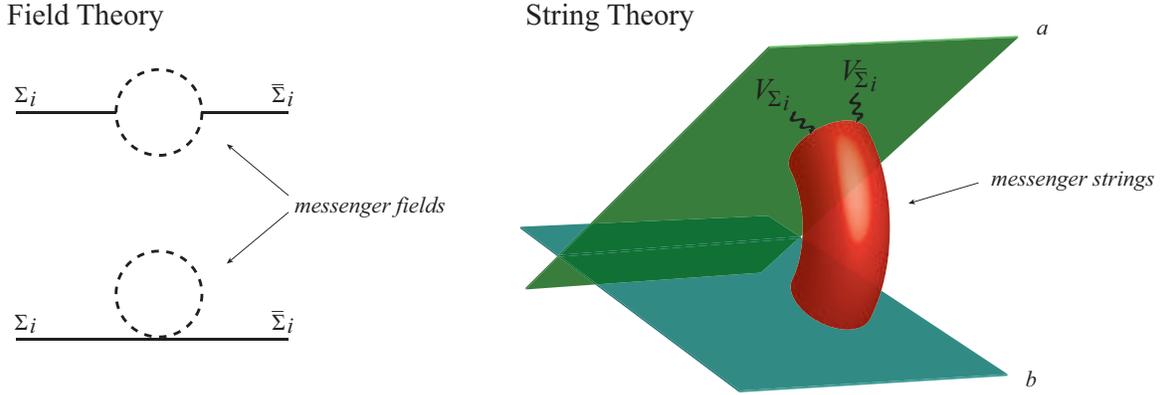,width=15.3cm}
\caption{Equivalence of Field and String theory amplitudes. In field theory, supersymmetry is broken at tree level for the messengers running in the loops and it is communicated to $\Sigma_i,~\bar\Sigma_i$ at 1-loop. In string theory, the role of the messenger fields is played by strings streached between the two branes.}
\label{1loopmass_in_FT}
\end{center}
\end{figure}

As we have already mentioned, strings at the intersections feel the breaking of supersymmetry and they communicate it to the rest of the strings living on supersymmetric configurations.
The scheme is similar to the field theory where non-supersymmetric messenger field generate mass split at 1-loop to other supersymmetric fields.
In the string theory, the role of the messengers is played by the strings at the angles running in a loop (see figure \ref{1loopmass_in_FT}).

\subsubsection{The ${\cal N}  \approx 1$ case}

In such configuration D-branes intersect at all tori.  
The corresponding surface with boundaries is the annulus\footnote{In principle, we should also consider the M\"obius strip. However, we dont expect any effect on the supersymmetry breaking by this amplitude since there is no change of the angle between the OD-branes and the orientifold planes.} with the two VOs are inserted at the same boundary.
The corresponding diagrams are:
\bea
{\cal A}_{\Sigma_3 \Sigma_3} &=& ig^2 \int^\infty_0 {dt} \int^{it/2}_0 {dz_1} \int^{it/2}_0 {dz_2}
\int {d^4p\over (2\pi)^4} Tr[V(k;z_1) V(k;z_2) e^{L_0}]\nn\\
{\cal A}_{\Sigma_3\bar \Sigma_3} &=& ig^2 \int^\infty_0 {dt} \int^{it/2}_0 {dz_1} \int^{it/2}_0 {dz_2}
\int {d^4p\over (2\pi)^4} Tr[V(k;z_1)\bar V(k;z_2) e^{L_0}]
\label{1-loop amps}
\eea
where the vertex operators for the adjoint scalars are:
\bea
V_{\Sigma_i}(k,z)&=&2 g ~ (i \partial Z^i+ \alpha ' (k\cdot \psi) \Psi^i) e^{i k\cdot X(z)}\nn\\
\bar V_{\bar \Sigma_i}(k,z)&=&2 g ~ (i \partial \bar Z^i - \alpha ' (k\cdot \psi) \bar \Psi^i) e^{-i k\cdot X(z)}
\eea
and $k_\mu$ the momenta four-vectors.

The traces in (\ref{1-loop amps}) run over all word-sheet fields living on the annulus which is stretched between the D-branes $a$, $b$. By $p^\mu$ we denote the momenta running in the loop. 
The integrals run over all possible positions of the VOs $z_1,~z_2\in [0,it/2)$ and the size of the annulus $t\in [0,\infty)$. Using translational invariance on the annulus we fix the second VO at zero: $z_2=0$.

The above amplitudes are zero on-shell if we enforce the conditions $k^2=0$. There is however a consistent off-shell extension which has given consistently the mass of bosons in other cases \cite{Poppitz:1998dj, Bain:2000fb, akr, pascalU1masses, Berg:2011ij} and we adopt it here.
We will impose these conditions only at the end of our calculations (after all integrations are performed).
The amplitudes are:
\bea
&&{\cal A}_{\Sigma_3 \bar \Sigma_3} = 
- {g^2 \over 16\pi^4 \alpha'^3} \int^\infty_0 {dt\over t^2} \int^{it/2}_0 {dz_1}~
\nn\\
&&~~~~~~~~~~~~~~~~~~~
{\cal Z}_{{\cal M}_4}^{\cal B} Z_{ghost}^{\cal B} 
{\cal Z}_{{\cal T}_1^2}^{{\cal B},\theta^1}
{\cal Z}_{{\cal T}_2^2}^{{\cal B},\theta^2}
{\cal Z}_{{\cal T}_3^2}^{{\cal B},\theta^3}
\nn\\
&&~~~~~~~~~~~~~~~~~~~
\times\sum_{ab} C[^a_b] 
{\cal Z}_{{\cal M}_4}^{\cal F}[^a_b] 
{\cal Z}_{ghost}^{\cal F}[^a_b] 
{\cal Z}_{{\cal T}_1^2}^{{\cal F},\theta^1}[^a_b]
{\cal Z}_{{\cal T}_2^2}^{{\cal F},\theta^2}[^a_b]
{\cal Z}_{{\cal T}_3^2}^{{\cal F},\theta^3}[^a_b]
\nn\\
&&~~~~~~~~~~~~~~~~~~~
\times 
\Big\langle  e^{i k\cdot X(z_1)} e^{-i k\cdot X(0)}\Big\rangle  
\nn\\
&&~~~~~~~~~~~~~~~~~~~
\times  \Big(\Big\langle \partial Z^3(z_1)
\partial \bar Z^3 (0)\Big\rangle 
+ \alpha'^2 k^2 \Big\langle \psi(z_1) \psi(0)
\Big\rangle[^a_b] \Big\langle 
\Psi^3(z_1) \bar \Psi^3(0) \Big\rangle[^a_b] \Big)
\eea
The correlation and the partition functions are given in the appendix. 
The sum over the spin structures is performed using the Riemann identity.
After several steps we get an expression given only in terms of the well known theta-function $\vartheta_1 (z, it/2)$:
\bea
{\cal A}_{\Sigma_3\bar\Sigma_3}&\approx& 
-{2ig^2{
{k^2}} \over 16\pi^4} I_{ab}
 \int^\infty_0 {dt\over t^2}
{\vartheta'_1 (0)^2 \over \h^6}
{{\vartheta_1((\theta_2 -\epsilon) it/2)}~
{\vartheta_1((\theta_3 -\epsilon) it/2)}\over
\vartheta_1(\theta_1 it/2)~\vartheta_1(\theta_2 it/2)~\vartheta_1(\theta_3 it/2)}
\\
&&~~~~~~~~~~~~\times \int^{it/2}_0 {dz_1} ~ 
 e^{i {
 {k^2}} \langle X(z_1)X(0)\rangle} 
 e^{2\p i z_1 \theta_1}
{\vartheta_1 (z_{1}+\epsilon it/2)~
\vartheta_1 (z_{1}+(\theta_1 -\epsilon) it/2)
\over  \vartheta_1 (z_{1})^2}~~~~\nn
\eea
where $I_{ab}$ is the intersection number of the two D-branes.

The last non-trivial step is to perform the integrals. Notice that the amplitude has an overall factor $k^2$. Thus, the only way to get a non-vanishing result on shell ($k^2=0$) is to find the integration limits that can provide a factor $1/k^2$. 
There are three different limits which we can generate a mass term for the above amplitude (figure \ref{Limits_Fig}): 
\begin{figure}[t]
\begin{center}
\epsfig{file=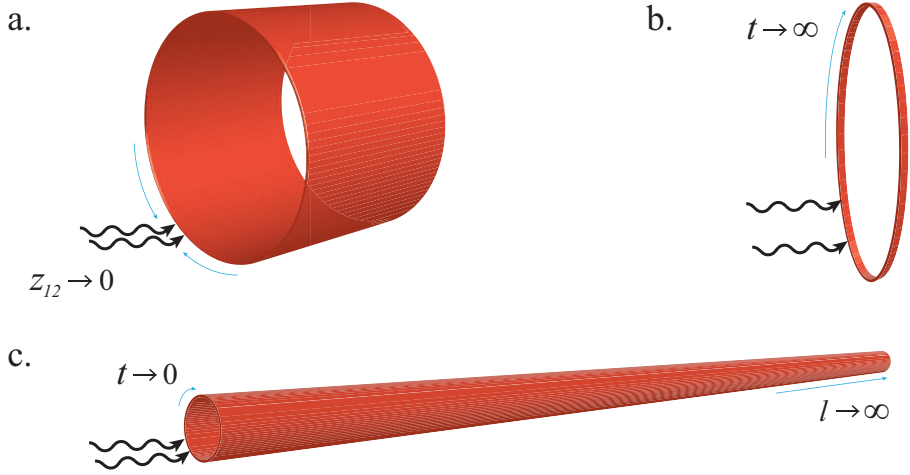,width=12cm}
\caption{The three different limits which can generate a mass term for 1-loop open string amplitudes.}
\label{Limits_Fig}
\end{center}
\end{figure}
\begin{itemize}
\item[a.] World-sheet poles coming from the limit $z_{12}\to 0$. Single poles give us momentum poles via:
\bea
{\cal A} \sim k^2\int {dz_1} ~ \left( {\vartheta_1(z_1)\over \vartheta_1'(0)}\right)^{-1-2 \alpha' k^2}\sim k^2\int {dz_1} ~ \left( z_1 \right)^{-1-2 \alpha' k^2} \to {1\over 2 \alpha '}
\eea
whereas double poles do not contribute as $k^2 \to 0$ due to analytic continuation in $k^2$.

In our case, there are poles both at $z_1 = z_2=0$ and $z_1 = it/2+z_2 = it/2$, and they cancel.

\item[b.] At the closed string UV limit (long strip limit $t\to \infty$). This is quite uncommon but might appear due to massless open strings running in the loop \cite{Conlon:2010xb}.

In our case, there are no long strip contributions.

\item[c.] At the closed string IR limit (long-tube limit $t\to 0$). This is due to the massless close-string exchange between the two annulus boundaries (notice the change of variables $l=1/t$) \cite{akr, pascalU1masses}. 
\bea
{\cal A} &\sim& k^2\int {dl} ~ e^{-k^2 \langle XX\rangle(z_1)}~\\
&& ~~\sim~
\left\{
\begin{array}{llllll}
\mbox{VOs on opposite boundaries} ~&~:~
k^2 \displaystyle \int_{a}^\infty dl ~e^{-\pi \alpha' k^2 l}
&~~\to~& \displaystyle {1\over \pi \alpha'}\\
\\
\mbox{VOs on the same boundary} ~&~:~
k^2\displaystyle \int^\infty_a {dl} ~(2\sin \pi x)^{-2\alpha' k^2}
&~~\to~&
\mbox{tadpoles}
\end{array}
\right.\nn~~~
\eea

Our case, fall in the second category and there are long tube contributions:
\bea
{\cal A}_{\Sigma_3\bar\Sigma_3} = - {ig^2\over 16 \pi^3 \alpha'} 
\Big({{\cal V}^1_a{\cal V}^1_b \over T^1_2}
{{\cal V}^2_a{\cal V}^2_b \over T^2_2}
{{\cal V}^3_a{\cal V}^3_b \over T^3_2}\Big)
\Big(-1 + \cos^2[\pi\theta_1] + \cos^2[\pi\theta_2]- \cos^2[\pi\theta_3] \Big)~~~~\nn
\eea
where ${\cal V}_{a(b)}^i$ the world-volume of the D-brane in the $a~(b)$ directions of the $i$th torus.
Also $T^i_2$ is the K\"ahler modulus of the $i$th torus.

Notice however, that this tadpole does not depend on the supersymmetry breaking parameter $\epsilon$. 
This result was expected also in supersymmetric frameworks and it is cancelled in all consistent models (without R-R and NS-NS tadpoles) by a similar tadpole which comes from the M\"obius strip amplitude \cite{Poppitz:1998dj, Bain:2000fb}. 

\end{itemize}
We can therefore conclude that there is no contribution to the masses of the adjoint scalars from ${\cal N}  \approx 1$ sectors.

\subsection{Parallel directions by brane displacement method.}

\begin{figure}[t]
\begin{center}
\epsfig{file=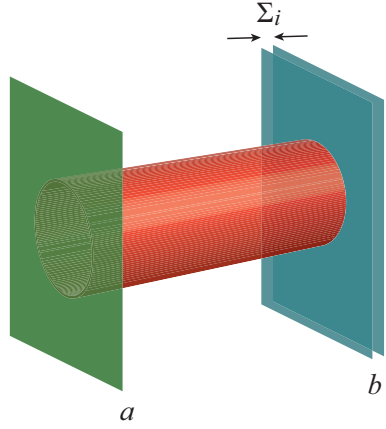,width=5cm}
\caption{Displacing the branes by $\Sigma_i$ in parallel directions.}
\label{Displacement_Fig}
\end{center}
\end{figure}

Next, we will evaluate the radiative corrections for the masses of adjoint scalars in parallel directions. Such scalars appear only in the ${\cal N}  \approx 2,4$ cases. We could perform a similar calculation to the above, but for sake of simplicity we will use a different approach: 
We will evaluate the 1-loop partition function at the presence of some displacements of the branes in the parallel directions (fig \ref{Displacement_Fig}). This will give us the potential for these dicplacements and we can easilly evaluate the induced 1-loop mass for these fields by taking the second derivative.

The 1-loop partition function is:
\bea
{\cal Z}(it/2)&~~\sim ~~&
{\cal Z}_{{\cal M}_4}^{\cal B} 
{\cal Z}_{ghost}^{\cal B} 
{\cal Z}_{{\cal T}_1^2}^{\cal B}
{\cal Z}_{{\cal T}_2^2}^{\cal B}
{\cal Z}_{{\cal T}_3^2}^{\cal B}~ \times ~
{\cal Z}_{{\cal M}_4}^{\cal F} 
{\cal Z}_{ghost}^{\cal F} 
{\cal Z}_{{\cal T}_1^2}^{\cal F}
{\cal Z}_{{\cal T}_2^2}^{\cal F}
{\cal Z}_{{\cal T}_3^2}^{\cal F}
\eea
A displacement of the branes by $\Sigma_i$ in parallel directions would only affect the bosonic partition function of this torus. 
\bea
{\cal Z}_{{\cal T}_i^2}^{{\cal B},||}(\Sigma^i_1,l^i+\Sigma^i_2)
={1 \over \eta^2 }\sum_{m_i,n_i}
e^{2\pi i \tau \big(\big(\Sigma^i_1+{m_i\over R_{1,i}}\big)^2+\big(l^i+\Sigma^i_2+{n_i R_{2,i}}\big)^2\big)} 
\eea
Next, we will consider separately the ${\cal N}  \approx 2,~4$ cases.

\subsubsection{The ${\cal N}  \approx 2$ case}

The total 1-loop partition function is given by
\bea
{\cal Z}(it/2)&\sim &
{\cal Z}_{{\cal M}_4}^{\cal B} 
{\cal Z}_{ghost}^{\cal B} 
{\cal Z}_{{\cal T}_1^2}^{\cal B,||}(l)
{\cal Z}_{{\cal T}_2^2}^{\cal B,\angle}(\theta_2)
{\cal Z}_{{\cal T}_3^2}^{\cal B,\angle}(\theta_3) \nn\\
&&
{\cal Z}_{{\cal M}_4}^{\cal F} 
{\cal Z}_{ghost}^{\cal F} 
{\cal Z}_{{\cal T}_1^2}^{\cal F,||}(l)
{\cal Z}_{{\cal T}_2^2}^{\cal F,\angle}(\theta_2)
{\cal Z}_{{\cal T}_3^2}^{\cal F,\angle}(\theta_3)
\eea
where we insert a displacement in the first torus. The potential for the ${\cal N}  \approx 2$ case is:
\bea
{\cal V}(\Sigma^1_1,\Sigma^1_2)
&=& 
{ i\over 32\pi^4 \alpha'}  
\int^\infty_0 {dt\over t^3} 
{\cal Z}_{{\cal M}_4}^{\cal B} 
{\cal Z}_{ghost}^{\cal B} 
{\cal Z}_{{\cal T}_1^2}^{{\cal B},||}(\Sigma^1_1,l^1+\Sigma^1_2)
{\cal Z}_{{\cal T}_2^2}^{{\cal B},\theta^2}
{\cal Z}_{{\cal T}_3^2}^{{\cal B},\theta^3}
\nn\\
&&~~~~~~~~~~~~~~~~~
\times\sum_{ab} C[^a_b] 
{\cal Z}_{{\cal M}_4}^{\cal F}[^a_b] 
{\cal Z}_{ghost}^{\cal F}[^a_b] 
{\cal Z}_{{\cal T}_1^2}^{{\cal F},||}[^a_b]
{\cal Z}_{{\cal T}_2^2}^{{\cal F},\theta^2}[^a_b]
{\cal Z}_{{\cal T}_3^2}^{{\cal F},\theta^3}[^a_b]
\eea
By taking derivatives on $\Sigma_i$'s and letting afterwards $\Sigma_i\to0$ we get the tadpoles:
\bea
\ &&		{\cal V}^{(0,1)} ~\sim~- 32 \pi^2\varepsilon^2 \displaystyle \sum_{m,n}  \frac{l+nR_{2,1}}{(m R_{1,1})^2+\left(l+nR_{2,1}\right)^2 }~~\neq~~ 0\\
\ &&		{\cal V}^{(1,0)}~\sim~  - 32 \pi^2\varepsilon^2 \displaystyle \sum_{m,n}  \frac{mR_{1,1}}{(m R_{1,1})^2+\left(l+nR_{2,1}\right)^2 }~~\rightarrow~~0
\eea
where by ${\cal V}^{(i,j)}$ we denote the $i$th, $j$th derivatives of the potential over $\Sigma_1,~\Sigma_2$ respectively.

Notice that the ${\cal V}^{(1,0)}$ is an odd function on $m$ and consequently vanishes running over all integers from $-\infty$ to $\infty$. The tadpole from ${\cal V}^{(0,1)}$ can also be eliminated if we add another secluded brane at distance $-l$.

In order to get the mass-matrix for the adjoint fields we evaluate second derivatives on $\Sigma_i$'s:
\bea
\ &&		{\cal V}^{(2,0)}~\sim ~~~ 32\pi^2\varepsilon^2\displaystyle \sum_{m,n}  \frac{(m R_{1,1})^2+\left(l+nR_{2,1}\right)^2}{\left[(m R_{1,1})^2+\left(l+nR_{2,1}\right)^2\right]^2 } ~~\leq~~0\\
\ &&		{\cal V}^{(0,2)}~\sim -32\pi^2\varepsilon^2\displaystyle \sum_{m,n}  \frac{(m R_{1,1})^2-\left(l+nR_{2,1}\right)^2}{\left[(m R_{1,1})^2+\left(l+nR_{2,1}\right)^2\right]^2 } ~~\geq~~0\\
\ &&		{\cal V}^{(1,1)}~\sim ~~~64\pi^2\varepsilon^2\displaystyle \sum_{m,n}  \frac{(m R_{1,1})\left(l+nR_{2,1}\right)}{\left[(m R_{1,1})^2+\left(l+nR_{2,1}\right)^2\right]^2 }~\rightarrow~~ 0
\eea
However, the mass-matrix for the adjoint scalars is traceless (since ${\cal V}^{(2,0)}=-{\cal V}^{(0,2)}$) denoting the presence of two opposite-sign entries. This fact leads to tachyonic states. Their presence cannot be annihilated with the addition of an image brane.

\subsubsection{The ${\cal N}  \approx 4$ case}

In this case the partition function is:%
\bea
{\cal V}(it/2)&\sim &
{\cal Z}_{{\cal M}_4}^{\cal B} 
{\cal Z}_{ghost}^{\cal B} 
{\cal Z}_{{\cal T}_1^2}^{\cal B,||}(l_1)
{\cal Z}_{{\cal T}_2^2}^{\cal B,||}(l_2)
{\cal Z}_{{\cal T}_3^2}^{\cal B,\angle}(\epsilon) \nn\\
&&
{\cal Z}_{{\cal M}_4}^{\cal F} 
{\cal Z}_{ghost}^{\cal F} 
{\cal Z}_{{\cal T}_1^2}^{\cal F,||}(l_1)
{\cal Z}_{{\cal T}_2^2}^{\cal F,||}(l_2)
{\cal Z}_{{\cal T}_3^2}^{\cal F,\angle}(\epsilon)
\eea
We insert a displacement in both tori ${\cal T}^2_1,~{\cal T}^2_2$.
The potential for the ${\cal N}  \approx 4$ case reads:
\bea
{\cal V}(\Sigma_{1,i},~\Sigma_{2,i}) \sim -4\pi^2 \varepsilon ^3~
\bigg(\sum_{i=1,2}\left(({ \Sigma_{1,i}+ \tilde n_i R_{1,i}})^2+(\Sigma_{2,i}+ l_i+{n_i R_{2,i}})^2\right)\bigg)^{-1}
\eea
Taking derivatives on $\Sigma_{1,1},~\Sigma_{2,1},~ \Sigma_{2,1},~ \Sigma_{2,2}$ and setting all $\Sigma_{i,j}\to0$ we get the tadpoles:
\bea
{\cal V}^{(1,0,0,0)}
&~\sim~& {8\pi^2 |\varepsilon |^3 }~
\sum_{\tilde n,n} {\tilde n_1 R_{1,1}
\over 
(\sum_{i}\left(({\tilde n_i R_{1,i}})^2+(l_i+{n_i R_{2,i}})^2\right))^2}~~\to~~0\\
{\cal V}^{(0,1,0,0)}
&~\sim~& {8\pi^2 |\varepsilon |^3 }~
\sum_{\tilde n,n} {l_1+  n_2 R_{2,1}
\over 
(\sum_{i}\left(({\tilde n_i R_{1,i}})^2+(l_i+{n_i R_{2,i}})^2\right))^2}~~\neq~~~0\\
{\cal V}^{(0,0,1,0)}
&~\sim~& {8\pi^2 |\varepsilon |^3 }~
\sum_{\tilde n,n} {\tilde n_2 R_{1,2}
\over 
(\sum_{i}\left(({\tilde n_i R_{1,i}})^2+(l_i+{n_i R_{2,i}})^2\right))^2}~~\to~~0\\
{\cal V}^{(0,0,0,1)}
&~\sim~& {8\pi^2 |\varepsilon |^3 }~
\sum_{\tilde n,n} {l_1+  n_2 R_{2,2}
\over 
(\sum_{i}\left(({\tilde n_i R_{1,i}})^2+(l_i+{n_i R_{2,i}})^2\right))^2}~~\neq~~~0
\eea
The non-vanishing tadpoles can be cancelled by properly choosing image branes (since the tadpoles are odd on the distances $l_i$ we can put image branes on distance $-l_i$).

In order to get the mass-matrix we evaluate:
\bea
\begin{array}{rrrrrrr}
{\cal V}^{(2,0,0,0)}
&~\sim~&~16 i \pi^2 |\varepsilon|^3
\displaystyle \sum_{\tilde n,n} {-4( \tilde n_1 R_{1,1})^2 + S[\tilde n,n]
\over 
S[\tilde n,n]^3}
~~~&~~~ \neq ~~ 0\\
{\cal V}^{(1,1,0,0)}
&~\sim~&~16 i \pi^2 |\varepsilon|^3
\displaystyle \sum_{\tilde n,n} {4 ( \tilde n_1 R_{1,1})(l_1 + n_1 R_{2,1})
\over 
S[\tilde n,n]^3}&
\to ~~ 0\\
{\cal V}^{(1,0,1,0)}
&~\sim~&~16 i \pi^2 |\varepsilon|^3
\displaystyle \sum_{\tilde n,n} {4 ( \tilde n_1 R_{1,1})( \tilde n_2 R_{1,2})
\over 
S[\tilde n,n]^3}&
\to~~0\\
%
%
{\cal V}^{(1,0,0,1)}
&~\sim~&~16 i \pi^2 |\varepsilon|^3
\displaystyle \sum_{\tilde n,n} {4 ( \tilde n_1 R_{1,1})(l_2 + n_2 R_{2,2})
\over 
S[\tilde n,n]^3}&
\to~~ 0\\
{\cal V}^{(0,2,0,0)}
&~\sim~&~16 i \pi^2 |\varepsilon|^3
\displaystyle \sum_{\tilde n,n}{-4 (l_1 + n_1 R_{2,1})^2 + S[\tilde n,n]
\over 
S[\tilde n,n]^3}&
\neq~~0\\
%
%
{\cal V}^{(0,1,1,0)}
&~\sim~&~16 i \pi^2 |\varepsilon|^3
\displaystyle \sum_{\tilde n,n} {4 (l_1 + n_1 R_{2,1})( \tilde n_2 R_{1,2})
\over 
S[\tilde n,n]^3}&
\to~~0\\
%
{\cal V}^{(0,1,0,1)}
&~\sim~&~16 i \pi^2 |\varepsilon|^3
\sum_{\tilde n,n} {4 (l_1+ \tilde n_1 R_{1,1})( l_2 + n_2 R_{2,2})
\over 
S[\tilde n,n]^3}&
\neq~~0\\
{\cal V}^{(0,0,2,0)}
&~\sim~&~16 i \pi^2 |\varepsilon|^3
\displaystyle \sum_{\tilde n,n} {-4( \tilde n_2 R_{1,2})^2 + S[\tilde n,n]
\over 
S[\tilde n,n]^3}&
\neq~~0\\
{\cal V}^{(0,0,1,1)}
&~\sim~&~16 i \pi^2 |\varepsilon|^3
\displaystyle \sum_{\tilde n,n} {4 ( \tilde n_2 R_{1,2})(l_2 + n_2 R_{2,2})
\over 
S[\tilde n,n]^3}&
\to~~0\\
{\cal V}^{(0,0,0,2)}
&~\sim~&~16 i \pi^2 |\varepsilon|^3
\displaystyle \sum_{\tilde n,n}{-4 (l_2 + n_2 R_{2,2})^2 + S[\tilde n,n]
\over 
S[\tilde n,n]^3}&
\neq~~0
\end{array}
\eea
where $S[\tilde n,n]=(\tilde n_1 R_{1,1})^2 + (l_1 + n_1 R_{2,1})^2+( \tilde n_2 R_{1,2})^2 + (l_2 + n_2 R_{2,2})^2$.
Here again, all odd functions on the winding numbers vanish.
Schematically, the mass matrix is given by:
\bea
{\cal M}^2_{N\approx 4}~ \sim 
{|\epsilon|^3 g^2 |I_{ab}| \over
32 \pi^2 \alpha'}{\tiny
\left(~ \begin{array}{cccc} 
A_{1,2}^2 + A_{2,1}^2 + A_{2,2}^2 - 3 A_{1,1}^2 &0 & 0 & 0 \\
0 & A_{1,1}^2 + A_{2,1}^2 + A_{2,2}^2 - 3 A_{1,2}^2 &0 & - A_{1,2} A_{2,2}\\
0 & 0 & A_{1,1}^2 + A_{1,2}^2 + A_{2,2}^2 - 3 A_{2,1}^2 &0 \\
0 & - A_{1,2} A_{2,2} & 0 & A_{1,1}^2 + A_{1,2}^2 + A_{2,1}^2- 3 A_{2,2}^2 \end{array} ~ \right)}
~~~~~~~~~\eea
and again it is traceless. Therefore, there is at least one tachyonic state.

\section{Conclusions}

We considered breaking of supersymmetry in intersecting D-brane configurations by
slight deviation of the angles from their supersymmetric values.
We computed the masses generated by radiative corrections for the adjoint scalars on the
brane world-volumes.
  
In the open string channel, the string two-point function receives contributions only
from the infrared (${\cal N}  \approx 2,~4$) and the ultraviolet limits (${\cal N}  \approx 1$).
The latter is due to tree-level closed string uncanceled tadpoles which will be eliminated in all consistent models (without R-R, NS-NS tedpoles).

On the other hand, the infrared region (${\cal N}  \approx 2,~4$) reproduces the one-loop mediation of
supersymmetry breaking in the effective gauge theory, via messengers and their
Kaluza-Klein excitations.

\section*{Acknowledgments}

We would like to thank Carlo Angelantonj, Marcus Berg, Emilian Dudas, Eran Palti and Robert Richter for interesting discussions.
PA was supported by FWF P22000.%
MDG was supported by SFB grant 676 and ERC advanced grant 226371.
This work was supported in part by the European Commission under ERC Advanced Grant 226371 and contract PITN-GA-2009-237920.

We would also like to thank the organizers of Corfu Summer Institute 2011 School and Workshops on Elementary Particle Physics and Gravity for allowing us to present our work.

\appendix
\renewcommand{\theequation}{\thesection.\arabic{equation}}
\addcontentsline{toc}{section}{Appendices}
\section*{APPENDIX}
\bigskip\appendix

\section{Theta functions and modular invariance} \label{thetas}

In this short appendix, we establish our conventions for the modular
theta functions and list a few useful properties. The theta functions
are defined by:
\begin{equation}
\vartheta[ ^a_b](z| \tau) = \sum_{n =
-\infty}^{\infty} e^{\pi i (n+a)^2 \tau + 2 \pi i (n+a)(z+b)}
\end{equation}
where $\tau$ is the (complex) modular parameter of the torus, not to
be confused with the world-sheet coordinate used in the text. On the
cylinder, this parameter is purely imaginary and in the main text we
use the definition $\tau = i t/2$.
Alternatively, the theta functions can be defined as an infinite
product:
\begin{equation}
\vartheta[ ^a_b](z| \tau) = e^{2 \pi i a
(b+z)} q^{\frac{1}{2}a^2} \prod_{n \geq 1} (1 + q^{n+a - \frac{1}{2}}
e^{2 \pi i (b + z)}) (1 + q^{n - a - \frac{1}{2}} e^{-2 \pi i (b +
z)}) (1 - q^n) \label{theta_defined}
\end{equation} 
where $q = e^{2 \pi i \tau}$.
Defining
$\vartheta_1(z) \equiv \vartheta
[ ^{1/2}_{1/2}](z| \tau)~,~~
\vartheta_2(z) \equiv \vartheta
[ ^{1/2}_{~0}](z| \tau)~,~~
\vartheta_3(z) \equiv \vartheta
[ ^{0}_{0}](z| \tau)~,~~
\vartheta_4(z) \equiv \vartheta
[ ^{~0}_{1/2}](z| \tau)$.
In particular $\vartheta_1(z) = z \vartheta'_1(0) + \cdots \ $
a fact that was used repeatedly in the main text.

The Dedekind eta function is
$\eta(\tau) = q^{\frac{1}{24}} \prod_{n \geq 1} (1- q^n)$
and it is related to the function $\vartheta_1(z)$ by the simple identity
\begin{equation}
\vartheta'_1(0) = - 2 \pi \eta(\tau)^3
\end{equation}

Finally, the theta functions satisfy the following Riemann identity:
\begin{equation}
\sum_{a,b}C_{a,b}\vartheta[ ^a_b](z_1)
\vartheta[ ^a_b] (z_2)
\vartheta[ ^a_b](z_3) 
\vartheta[ ^a_b](z_4)=2
\vartheta_1(z'_1)
\vartheta_1(z'_2) 
\vartheta_1(z'_3) 
\vartheta_1(z'_4)
\label{riemann}
\end{equation}
with
$ z'_1= \frac {1}{2}(z_1+z_2+z_3+z_4),~ z'_2= \frac
{1}{2}(z_1+z_2-z_3-z_4),~ z'_3 = \frac {1}{2}
(z_1-z_2+z_3-z_4),~ z'_4= \frac {1}{2}(z_1-z_2-z_3+z_4).$

\section{Partition functions} \label{PF}

The partition functions of the bosonic and fermionic modes are:
\bea
\begin{array}{lllllll}
{\bf Bosonic} && {\bf Fermionic}\\ 
\mathcal{Z}_{{\cal M}_4}^{\cal B} ={1\over \h^4} &~~~~~~~~~~~~~~~&
\mathcal{Z}_{{\cal M}_4}^{\cal F} ={\vartheta[_{a}^{b}]^2 \over \h^2} \\
\mathcal{Z}_{ghost}^{\cal B}  ={\h^2} &&
\mathcal{Z}_{ghost}^{\cal F}  ={\h \over \vartheta[_{a}^{b}]} \\
\mathcal{Z}_{{\cal T}_i^2}^{{\cal B},||}(l_i)= {\Lambda_{m_i,n_i} (l_i) \over \h^2 } =
{1 \over \h^2 }\sum_{m_i,n_i}
e^{2\pi i \tau \big[({m_i\over R_{1,i}})^2+(l_i+{n_i R_{2,i}})^2\big]} &&
\mathcal{Z}_{{\cal T}_i^2}^{{\cal F},||}= {\vartheta[_{a}^{b}] \over \h}\\
\mathcal{Z}_{{\cal T}_i^2}^{{\cal B},\angle} (\a)=  {\h \over \vartheta[_{~~1/2}^{1/2+\a}]} &&
\mathcal{Z}_{{\cal T}_i^2}^{{\cal F},\angle} (\a)= {\vartheta[_{~a}^{b+\a}] \over \h}~~~~~~~~
\end{array}~~~~
\eea
where we have omitted the argument of the theta-function which are evaluated in $\tau=it/2$, $\nu=0$. In addition we have introduced twisted theta function and we have indicated with $ab$ the spin structures.

We also use the Poisson resummation formula:
\be
\sum_{n\in Z}e^{-\pi a n^2+\pi b n}={1\over \sqrt{a}}\sum_{n\in Z}
e^{-{\pi\over a}\left(n+i{b\over 2}\right)^2}
\label{PoissonResum} \ee
in order to T-dualize the longitudinal directions of the brane $X_i$.

\section{Correlation functions} \label{CF}

The untwisted correlators on the torus are:
\bea
&& \langle X(z_1) X(z_2) \rangle_{\cal T} = -{1\over 4} \log \left| {\vartheta_1(z_{12}| \tau) \over 
\vartheta'_1(0| \tau)}  \right|^2 +{\p \Im^2(z_{12}) \over 2 \tau_2}
\label{UntwistedXX}\\
&&\langle \partial Z(z_1) Z(z_2) \rangle_{\cal T} = -{1\over 4} {\vartheta_1'(z_{12}| \tau) \over 
\vartheta_1(z_{12}| \tau)}+{\p \over 2\tau_2} \partial_{z_{12}}{\Im^2(z_{12})}
\label{UntwistedXdX}\\
&&\langle \y(z_1) \y(z_2) \rangle _{{\cal T}} = {i\over 2}  {\vartheta[_{a}^{b}] (z_{12}| \tau) \vartheta'_1 (0| \tau) 
\over  \vartheta_1 (z_{12}| \tau)  \vartheta[_{a}^{b}] (0| \tau)} 
\label{UnwistedPSIPSI}\eea
The twisted correlators:
\bea
&& \langle Z(z_1)\partial \bar Z (z_2)\rangle_{{\cal T},\alpha} = -\frac 12 {\vartheta_{1}^{\alpha}(z_{12}) \vartheta_{1}'(0) \over \vartheta_1^\alpha(0) \vartheta_1(z_{12})} \label{TwistedXX}\\
&&\langle \y(z_1) \y(z_2) \rangle_{{\cal T},\alpha} = {i\over 2}  {\vartheta[_{a}^{b}]^\a (z_{12}| \tau) \vartheta'_1 (0| \tau) 
\over  \vartheta_1 (z_{12}| \tau)  \vartheta^\a[_{a}^{b}] (0| \tau)}
\label{TwistedPSIPSI}\eea
Notice that all correlation functions are periodic on the torus: $x_{12}\to x_{12}+ 1$ and  $x_{12}\to x_{12}+ \tau$.

In order to define the correlators on the annulus and M\"obius strip we use the involution: $I_{\cal A} =I_{\cal M} = 1-\bar z$:
\bea
\langle X(z_1) X(z_2) \rangle_{\s}&=&{1\over 2} 
\Big( 
\langle X(z_1) X(z_2) \rangle_{\cal T}  
+\langle X(z_1) X(I_{\s}(z_2)) \rangle_{\cal T} \nn\\
&&~~~+\langle X(I_{\s}(z_1)) X(z_2) \rangle_{\cal T} 
+\langle X(I_{\s}(z_1)) X(I_{\s}(z_2)) \rangle_{\cal T} \Big)
\eea
In our case, $z_1=i v,~z_2=0,~\tau=i t/2$ that gives $
\langle X(z_1) X(z_2) \rangle_{\s}
=2\langle X(z_1) X(z_2) \rangle_{\cal T}$ and the untwisted correlators:
\bea
&& \langle X(z_{12}) X(0) \rangle_{\cal A} = -\log {\vartheta_1(z_{12}| \tau) \over 
\vartheta'_1(0| \tau)} +{2\p z_{12}^2 \over t} -{\p i \over 2}
\label{UntwistedXX1}\\
&&\langle \partial X(z_{12}) X(0) \rangle_{\cal A}
= -{\vartheta_1'(z_{12}| \tau) \over 
\vartheta_1(z_{12}| \tau)} +{4\p z_{12} \over t}\label{UntwistedXdX2}\\
&&\langle \y(z_{12}) \y(0) \rangle _{{\cal A}} = {i}  {\vartheta_{ab} (z_{12}| \tau) \vartheta'_1 (0| \tau) 
\over  \vartheta_1 (z_{12}| \tau)  \vartheta_{ab} (0| \tau)} 
\label{UnwistedPSIPSI1}\eea
The twisted correlators:
\bea
&& \langle Z(z_1)\partial \bar Z (z_2)\rangle_{{\cal A},\alpha} = - {\vartheta_{1}^{\alpha}(z_{12}) \vartheta_{1}'(0) \over \vartheta_1^\alpha(0) \vartheta_1(z_{12})} \label{TwistedXX1}\\
&&\langle \y(z_1) \y(z_2) \rangle_{{\cal A},\alpha} = {i}  {\vartheta_{ab}^\a (z_{12}| \tau) \vartheta'_1 (0| \tau) \over  \vartheta_1 (z_{12}| \tau)  \vartheta^\a_{ab} (0| \tau)}
\label{TwistedPSIPSI1}\eea

\end{document}